# Compact high-resolution spectrographs for large and extremely large telescopes: using the diffraction limit

J. Gordon Robertson*[a] and Joss Bland-Hawthorn[a]

[a]Sydney Institute for Astronomy, School of Physics, University of Sydney, NSW 2006, Australia

## ABSTRACT

As telescopes get larger, the size of a seeing-limited spectrograph for a given resolving power becomes larger also, and for ELTs the size will be so great that high resolution instruments of simple design will be infeasible. Solutions include adaptive optics (but not providing full correction for short wavelengths) or image slicers (which give feasible but still large instruments). Here we develop the solution proposed by Bland-Hawthorn and Horton: the use of diffraction-limited spectrographs which are compact even for high resolving power. Their use is made possible by the photonic lantern, which splits a multi-mode optical fiber into a number of single-mode fibers. We describe preliminary designs for such spectrographs, at a resolving power of $R \sim 50,000$. While they are small and use relatively simple optics, the challenges are to accommodate the longest possible fiber slit (hence maximum number of single-mode fibers in one spectrograph) and to accept the beam from each fiber at a focal ratio considerably faster than for most spectrograph collimators, while maintaining diffraction-limited imaging quality. It is possible to obtain excellent performance despite these challenges. We also briefly consider the number of such spectrographs required, which can be reduced by full or partial adaptive optics correction, and/or moving towards longer wavelengths.

# 1. INTRODUCTION

The invention of the photonic lantern[1,2], which splits a multi-mode optical fiber into the appropriate number of single-mode fibers, brings with it the exciting possibility of developing compact high-resolution spectrographs which can be fed by almost any light source. Since a single-mode fiber (SMF) delivers a diffraction-limited input to the spectrograph, one can expect to obtain high spectral resolution without the huge optics which are the hallmark of seeing-limited spectrographs at large astronomical telescopes. This is the concept known as PIMMS#0[3,4]. It is of interest both for the visible spectrum (where adaptive optics has limited ability to narrow the image at the spectrograph slit) and the near-IR (where the single-mode fibers could incorporate fiber-Bragg gratings to suppress the atmospheric OH emission[5]), and also for laboratory instruments as well as at astronomical observatories. It could have particular application for remote sensing, planetary rovers, and space telescopes.

A further important development now under way is the replacement of individual SMFs by waveguides etched (by laser exposure) into a solid block. In this way the 'long-slit' array of hundreds (or thousands) of individual single-mode guides becomes feasible in a single integrated-optics unit.

However, design of a suitable high-resolution spectrograph to be fed by an array of single-mode fibers or waveguides is not a trivial exercise. The spectrograph optics must be close to diffraction-limited in order to avoid degrading the PSF significantly. As in any spectrograph design, we seek to obtain the maximum slit length (*i.e.* number of SMF feeds) and the maximum wavelength range on the detector.

A further problem specific to this application is that the beam from the single-mode guides is typically quite fast, *e.g.* a Gaussian beam of NA = 0.11 which means $f/4.5$ to the $1/e^2$ intensity points but about $f/3.4$ to avoid unacceptable truncation of the Gaussian beam. Hence the collimator must accept faster beams than those of a typical astronomical spectrograph.

Here we make a preliminary study of the visible-light version. We begin by studying relevant properties of Gaussian beams, as radiated by single-mode waveguides.

*Gordon.Robertson@sydney.edu.au; phone +61 2 9351 2825; fax +61 2 9351 7726

## 2. GAUSSIAN BEAM PROPERTIES

### 2.1 Beam width

Since a Gaussian beam has no 'edge', its width is specified to the points at which the intensity falls to $1/e^2$ of the central peak intensity. Thus the specification NA = 0.11 means sin θ = 0.11 where θ is the angle between the axis of the beam and the $1/e^2$ point. For NA = 0.11 we have θ = 6.315°.

A Gaussian drops to $1/e^2$ of its central intensity at points ±2σ from the peak location. The Full Width to Half Maximum (FWHM) of a Gaussian is 2.355σ, thus the relationship of the $1/e^2$ diameter to the FWHM is

$$d_{1/e^2} = \frac{4}{2.355} \times \text{FWHM} = 1.699 \times \text{FWHM} \,. \tag{1}$$

### 2.2 Beam Truncation

When a Gaussian beam is truncated by a finite-aperture optical element, several effects result:

a) Some throughput is lost due to light in the outer parts of the beam not reaching the image.

b) The FWHM of the PSF at the focal plane (*i.e.* the Gaussian beam waist) is broadened. This is significant even for moderate truncation, when the fractional power loss is small. The sensitivity of the width to loss of the extreme edges of the beam can be understood since it is the edges that are most critical in forming the width of the PSF (*i.e.* providing the high spatial frequencies).

c) For severe truncation, the form of the beam will depart from Gaussian. Eventually the image will approach an Airy function (*i.e.* with concentric rings) as the Gaussian is so severely truncated that it approaches a uniform aperture distribution.

The degree of truncation is measured by the truncation parameter $T$, defined such that

$$T = \frac{d_{1/e^2}}{D_{opt}} \tag{2}$$

where $D_{opt}$ is the diameter of the optical element limiting the beam.

ZEMAX[a] Physical Optics Propagation was used to study the effects of beam truncation. Figure 1 shows the results.

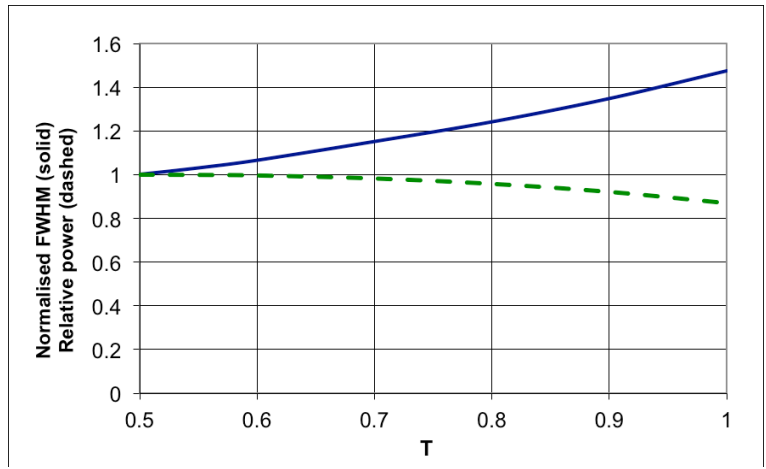

Figure 1. Solid blue curve: beam broadening expressed as the ratio of FWHM to the FWHM at T = 0.5 (where the effect of truncation is negligible). This has been computed using a Gaussian beam of NA=0.11, but the result is general. Dashed green curve: fractional power transmitted through the system as a function of truncation ratio.

[a] http://www.radiantzemax.com

A value of $T = 0.75$ represents a good compromise between minimising the amount of beam broadening (and power loss) while not making the optics unnecessarily large. At $T = 0.75$ the PSF is 20% broadened and about 3% of the power is lost, while the diameter of the optics is 1.33 times the nominal diameter to the $1/e^2$ points. In order to accommodate a Gaussian beam of NA = 0.11 out to $T = 0.75$, the optics must accept $f/3.4$. The optical systems shown below are in fact even faster than this, but $T = 0.75$ or $f/3.4$ is a good figure to keep in mind as representing the 'full' width of the Gaussian beam.

# 3.   SPECTRAL RESOLVING POWER

## 3.1  Introduction

The key aim of this project is to obtain high spectral resolution, so it is important to examine the factors which determine the diffraction-limited resolving power. We define:

| | |
|---|---|
| $\lambda$ | wavelength |
| $\lambda_c$ | center wavelength (to detector center) |
| $\Delta\lambda$ | wavelength interval |
| $D_{opt}$ | limiting diameter of optical elements |
| $f_{cam}$ | focal length of camera |
| $b$ | grating line spacing |
| $\theta_i$ | angle of incidence on grating |
| $\theta$ | angle of diffraction leaving grating |
| $m$ | diffraction order |
| $R$ | spectral resolving power $= \lambda/\Delta\lambda$ |

## 3.2  Uniformly illuminated pupil

We first treat the case of uniform amplitude across the aperture (often referred to as 'top-hat' illumination), which gives the familiar Airy function diffraction pattern. This will form a useful comparison with the later results for Gaussian beams. We consider the incident beam to be perfectly collimated, so our result will be the diffraction-limited resolution as opposed to a slit-limited resolution.

The diffraction angular spread from peak to the first null is given by the well-known $1.22\,\lambda/D_{opt}$ expression. Combining with the focal length of the camera we have:

$$\text{Linear size of resolution element} = \text{HWZI} = \frac{1.22\lambda f_{cam}}{D_{opt}} \qquad (3)$$

The half width to zero intensity (HWZI) is the same as the Rayleigh criterion resolution element, since two peaks separated by this amount are designated as being just resolved, *i.e.* the peak of one falls above the zero of the other. (This is the HWZI of a cross-cut rather than a projection of the 2D image on to 1D, but is sufficient for our purpose.)

We now find the linear displacement on the detector due to an increment $\Delta\lambda$ in wavelength.

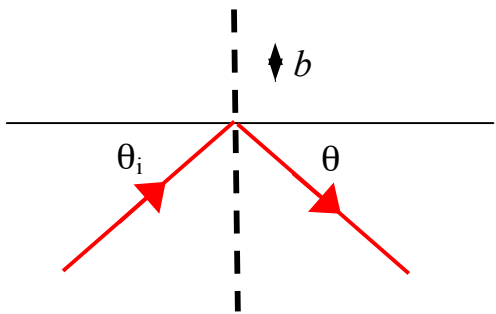

Figure 2. Illustrating the beam incident on the grating at angle $\theta_i$ and leaving at diffracted angle $\theta$. The line spacing in the grating is $b$.

The grating equation for the sign convention shown in Figure 2 is:

$$b \sin \theta_i + b \sin \theta = m\lambda .$$

(4)

Differentiating with respect to λ to obtain the angular dispersion and noting that $\theta_i$ is fixed and only θ varies with λ we obtain

$$\frac{\Delta \theta}{\Delta \lambda} = \frac{m}{b \cos \theta} .$$

(5)

We will assume the grating is operated at or close to Littrow configuration, *i.e.* $\theta_i = \theta$ (at the center wavelength). Hence the grating equation (4) becomes

$$2b \sin \theta_i = m\lambda_c .$$

(6)

Combining equations (5) and (6) and using the linear throw from the camera to the detector we obtain:

$$\text{linear displacement due to } \Delta\lambda \approx 2 f_{cam} \tan \theta_i \frac{\Delta \lambda}{\lambda_c}$$

(7)

We now equate the linear displacement due to Δλ (eqn 7) and the linear resolution element (eqn 3). By definition, the value of Δλ which results is the Rayleigh criterion limiting wavelength difference. Re-arranging and expressing the result as the spectral resolving power $R$ the result is:

$$R = \frac{\lambda_c}{\Delta \lambda} = 1.64 \tan \theta_i \frac{D_{opt}}{\lambda}$$

(8)

This can be compared with the well-known formula $R = mN$ where $N$ is the total number of (illuminated) lines in the grating. The resolving power in eqn (8) is lower than $mN$ by a factor 1.22, which is as expected because the circular aperture tapers down the effects of the grating lines of greatest separation, compared with $R = mN$ which assumes equal amplitude from all grating lines, *i.e.* a rectangular aperture – which is not used in practice.

### 3.3 Gaussian beam illumination

For a system which delivers a Gaussian beam we cannot directly use the Rayleigh criterion where the peak of one spectral resolution element's profile lies over the zero of the adjacent one. Instead we use the natural extension of this principle as follows. For a diffraction-limited rectangular spectrograph slit the intensity profile is the sinc$^2$ function. This does have zeros and the Rayleigh criterion can be applied. For the sum of two sinc$^2$ functions of equal magnitude separated at the Rayleigh criterion, the local minimum lying between the two peaks has a value of 81.1% of either peak. Now applying this to the sum of two offset but equal Gaussians, the local minimum between the two peaks has a value of 81% of the peak when the separation is 1.119 × FWHM. So two wavelengths will be considered to be just resolved when their separation on the detector is 1.119 times the FWHM of the (projected, monochromatic) diffraction pattern in the wavelength direction.

The spot size (*i.e.* beam waist, which is the same in cross-cut or projection) for a Gaussian beam at its focal plane is given by

$$\omega_0 = \frac{\lambda f}{\pi \omega}$$

(9)

where $\omega_0$ is the $1/e^2$ beam radius at the beam waist and ω is the corresponding radius on the lens of focal length $f$. This can be used to find the expected FWHM of the Gaussian PSF on the detector, then multiplying by the factor 1.119 to obtain the resolution element we have:

$$\text{linear resolution element} = 0.839 \frac{\lambda f_{cam}}{d_{1/e^2}}$$

(10)

The linear displacement due to a small wavelength shift is again given by eqn (7), so equating that and the value from eqn (10) gives the formula for spectral resolving power which is

$$R = 2.38 \tan \theta_i \, \frac{d_{1/e^2}}{\lambda} \qquad (11)$$

Comparing this result with eqn (8) for the resolving power with top-hat illumination, the Gaussian beam may appear to give higher resolution despite having its high spatial frequencies tapered down. But this impression is erroneous, because eqn (11) assumes the beam is untruncated, *i.e.* $D_{opt} > d_{1/e^2}$.

## 4. VERSION 1: DESIGN USING CATALOG OPTICS

In our first examination of the capabilities of a diffraction-limited spectrograph, we deal with an example of what can be achieved with catalog optics. We consider a spectrograph with the following properties:

- Centre wavelength 500 nm
- Spectral resolving power 50,000
- Input waveguides/SMFs each supply a Gaussian beam of NA = 0.11
- No constraint due to detector pixel size - *i.e.* pixels can be as small as required
- No constraint on overall detector size and hence number of pixels (the limit will be due to aberrations)

The key properties to be established are:

- A design which meets the target specifications
- The maximum 'slit-length' (hence number of waveguide cores) that can be accommodated before aberration becomes unacceptable.
- The maximum wavelength range that can be accepted on the detector before aberrations become too large.

### 4.1 Required system size

Whether seeing-limited or diffraction-limited, the size of a spectrograph increases as the required spectral resolving power is increased. But for a given resolving power the diffraction-limited instrument will be much smaller - which is the rationale behind the present concept.

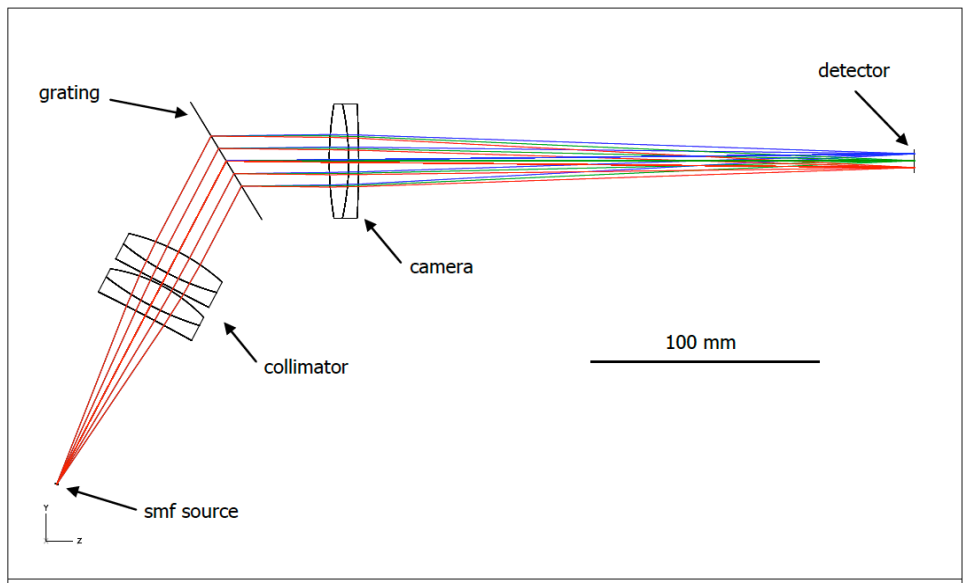

Figure 3. Layout diagram of the version 1 system as modeled. The rays shown emerge at NA = 0.11. The wavelengths shown are 495, 500 and 505 nm.

The starting point for the present design is to use eqn (11) to give an idea of the size of optics that will be needed. With R = 50,000 and $\lambda$ = 500 nm eqn (11) gives $\tan\theta_i \times d_{1/e2}$ = 10.5 mm. A higher dispersion grating could be used with smaller diameter optics. We expect to use Volume Phase Holographic (VPH) gratings and their efficiency is higher if the beam deviation $\theta$ is not too extreme. We initially chose $\theta \sim 27°$ which leads to $d_{1/e2} \sim 21$ mm. This results in conveniently sized optics.

Figure 3 shows the overall layout of the first system considered here.

## 4.2 Collimator

As noted above, the collimator will have to be at least as fast as $f/3.4$ if it is to avoid severe truncation of the NA = 0.11 Gaussian beam. Even this focal ratio produces 20% PSF broadening through truncation. Furthermore, the collimator must be even faster than this in order to accommodate a slit of some significant length. Focal ratios in this range are readily obtained in expensive multi-element custom-designed systems, but the essence of our concept is to use cheap components so that the systems can be replicated many times over. At the same time, the optical system must be close to diffraction-limited in order to avoid excessive aberration of the PSF. The required lens speed and aberration tolerance rule out the use of a single achromatic doublet as the collimator.

In this example system we use the CVI Melles Griot 'Fast Achromat' FAP-100.0-46.0 which consists of two cemented doublets with an air space between them, fixed together in a cylindrical mount. Using two doublets allows the system to accept a faster beam, and indeed these units are intended for tasks such as fiber coupling. This stock lens has focal length 100 mm and clear aperture 46 mm (*i.e.* $f/2.2$), and is designed for $\lambda$ 425 - 675 nm. Tests using ZEMAX showed that this lens system is not diffraction-limited when the whole aperture is illuminated, but it is close to diffraction-limited when illuminated with a beam of NA = 0.11.

## 4.3 Diffraction grating

Having chosen a collimator focal length of 100 mm, the collimated beam diameter is $d_{1/e2}$ = 22.1 mm. In this system significant resolving power is lost through aberration broadening and experimentation showed that a suitable value for the grating deviation is $\theta_i$ = 31.2°. From eqn (11) this gives $R$ = 63,700 but it will be reduced by aberrations to about 50,000.

As usual with VPH gratings we prefer to work in first order - for optimum grating efficiency and maximum Free Spectral Range between orders. Thus the grating equation (6) gives $b$ = 0.483 $\mu$m (2071 lines/mm). The grating has been modelled using a ZEMAX ideal transmission grating with this line spacing. The grating diameter is only about 55 mm and we expect that a VPH grating could readily be obtained with this or a closely similar line density and with negligible wavefront distortion.

## 4.4 Camera

An important issue for any diffraction-limited spectrograph is the small size of the final monochromatic image. For an Airy disc imaged at a focal ratio of $f_\#$ the spot has FWHM = 1.029 $\lambda f_\#$, while a focused Gaussian spot has $1/e^2$ diameter = 1.273 $\lambda f_{cam}/d_{1/e2}$, both of which will be just a few $\mu$m unless the camera has a very large focal ratio. This version 1 design assumes that detectors with the requisite small pixels will be available to suit any PSF spot size on the focal plane. However, while the use of a fast collimator was required by the waveguide NA, there is no reason to make the design more difficult by also using a fast camera. On the contrary, a long-focus camera will magnify the spots to a more readily sampled size. By using a long-focus camera we can expect diffraction-limited behavior (or close to it) from a single achromatic doublet. This design uses CVI Melles Griot LAO-250.0-50.0 which has focal length 250 mm and diameter 50 mm.

## 4.5 System Performance

Figure 4 shows the geometrical image quality delivered by the version 1 system. A slit length of ± 1 mm and a wavelength range of 492.5 – 505 nm can be accommodated while keeping the aberration spots within the notional Airy disc that would result from uniform illumination across a beam of NA = 0.11.

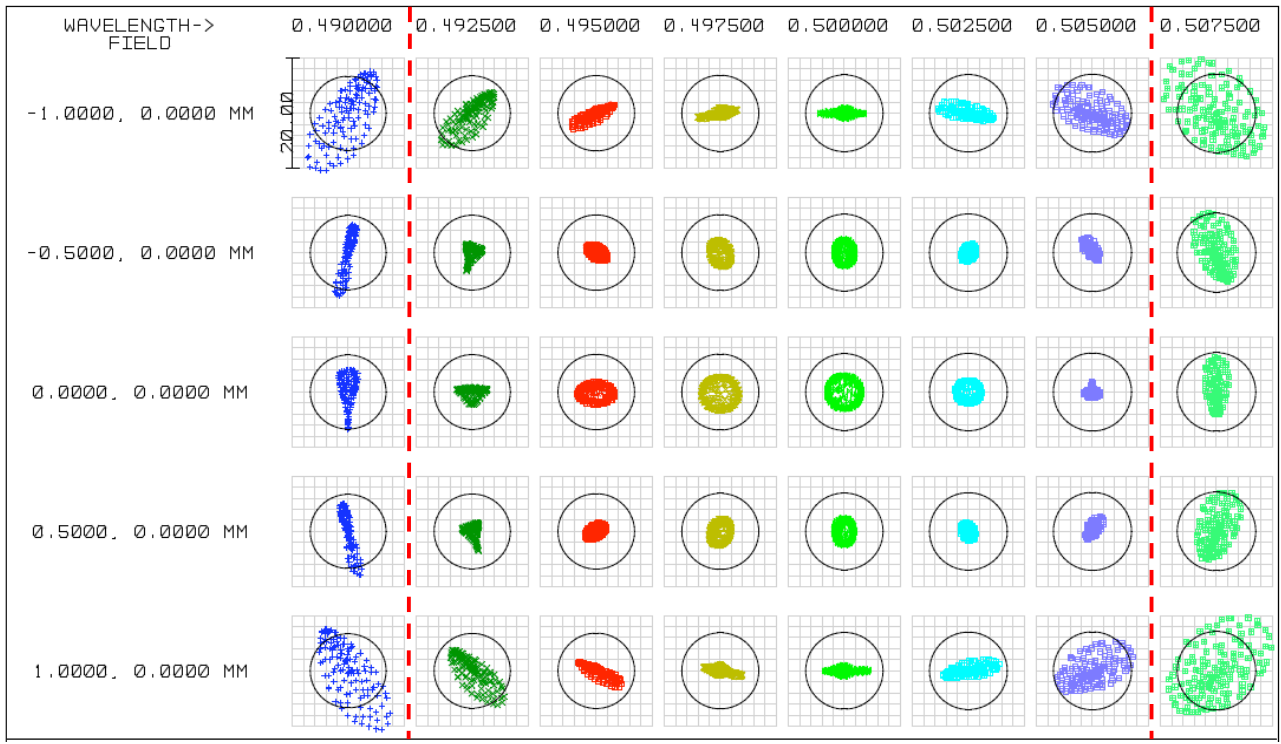

Figure 4. Spot diagrams from λ 490 nm to 507.5 nm and for feeds from -1mm to +1 mm along the slit. The illumination is NA=0.11 'top-hat'. The circles show the notional Airy ring radius, but diffraction is not taken into account in the spot diagrams. The box size is 20 μm. The images at the extreme wavelengths, outside the dashed red lines, are regarded as unacceptable and are not further considered.

Figure 5 shows the image quality when diffraction of the Gaussian beam is taken into account as well as aberrations, using the Physical Optics Propagation facility in ZEMAX. The 2-D images have been integrated to show the resulting 1-D profiles in both the wavelength and spatial directions. All profiles have clean Gaussian or quasi-Gaussian shapes, but some are significantly broadened by aberrations.

The theoretical diffraction-limited resolving power for an untruncated beam in this configuration (eqn 11) is 63,700. The actual spectral resolving powers obtained were 60,000 for the best profile (Figure 5 top left), 52,000 for the typical profile (not shown) and 44,900 for the worst profile (Figure 5 bottom left).

Assuming a 6 μm separation of waveguide cores along the pseudo-slit, this system would accommodate 333 separate single-mode inputs. It produces a wavelength range of about 1250 spectral resolution elements (at least 2500 wavelength pixels).

### 4.6 Version 1 - review

The version 1 design, using off-the-shelf components, demonstrates the principle of obtaining high resolving power from a compact spectrograph. But its resolving power varies with wavelength and slit position, due to significant aberrations. This system is not able to make best use of the diffraction-limited single-mode input, because the general-purpose catalog optical components are not optimised for this application.

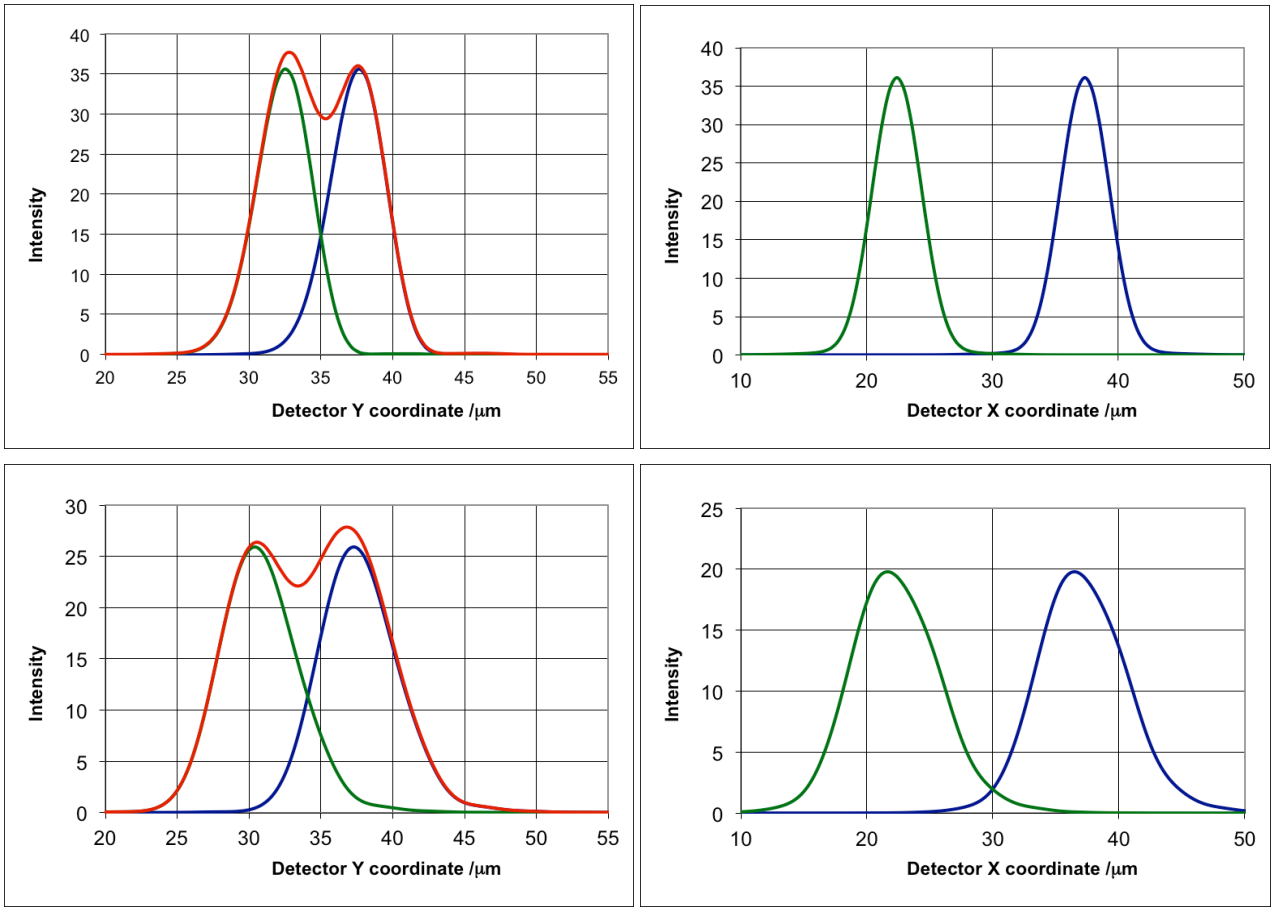

Figure 5. Projected profiles of the images on the detector, accounting for both diffraction and aberrations. Left-hand column: profiles integrated over the spatial direction and displayed along the wavelength axis. The same profile is shown twice, with an offset such that the two are at the resolution limit. The sum is shown in red. Right-hand column: profiles integrated over the wavelength direction and displayed along the spatial axis. Two profiles are shown, corresponding to two source waveguides separated by 6 μm. Top row: the best image quality (λ 505 nm, on axis). Bottom row: worst image quality (λ 505 nm, 1 mm off axis). The separations of the two profiles along the wavelength axis are 5.14 μm (top left) and 6.89 μm (bottom left).

## 5. VERSION 2: DESIGN USING CUSTOM OPTICS

The essence of the PIMMS#0 diffraction-limited spectrograph concept as applied to large or extremely large telescopes is that the instrument would be replicated many times - so instead of one enormous spectrograph (which in the case of an ELT might even be infeasible) we have an array of small cheap spectrographs that can still deliver high spectral resolution. Thus an actual instrument would still be a multi-million dollar investment, and so should not be restricted to catalog optics. The use of custom but nevertheless simple optics is clearly appropriate.

This section describes an example of what can be achieved by using custom optics. Compared with version 1, the system described in this section allows triple the slit length (now 6mm or 1000 cores at 6μm spacing), covers 2.2 × greater wavelength range and does so with significantly better image quality, hence better preserving the diffraction-limited resolution. It now feeds a 2×2 array of 4K×4K CCDs.

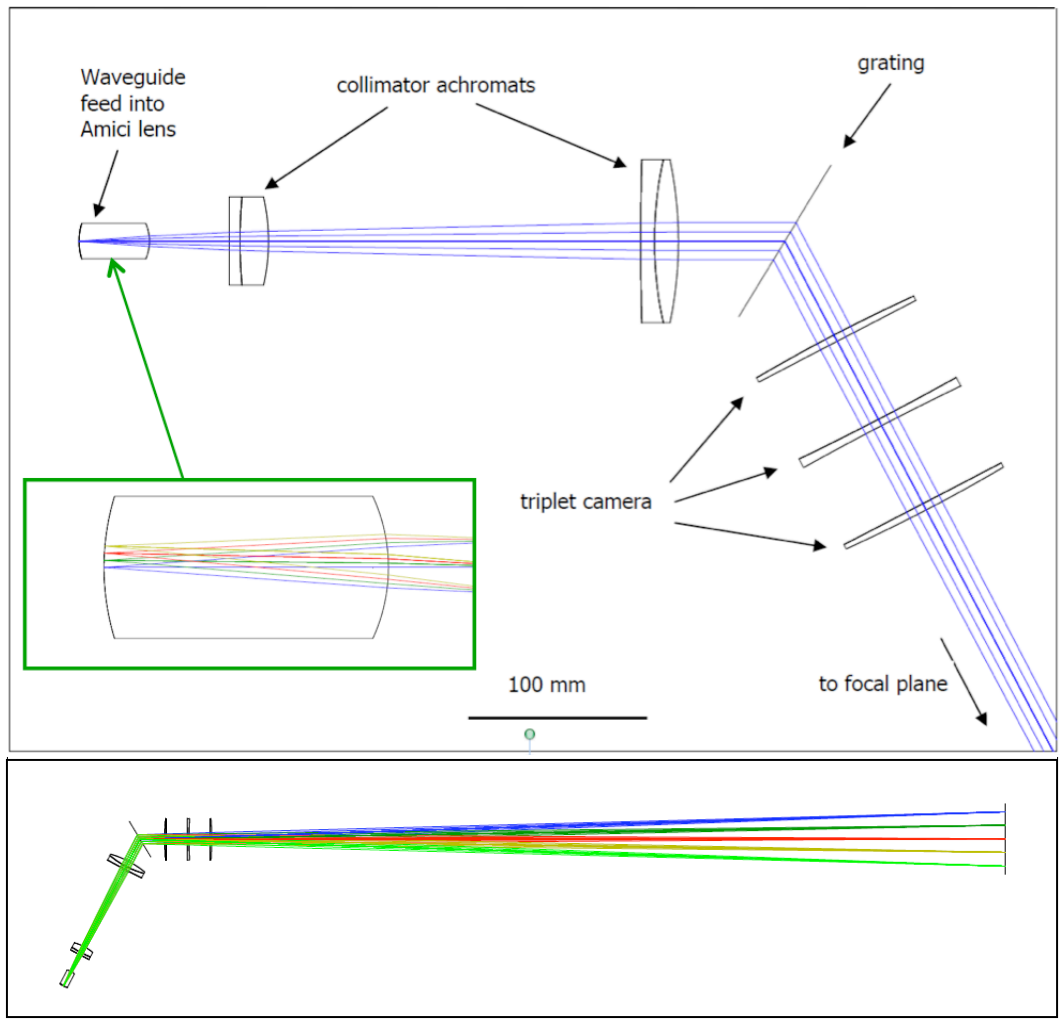

Figure 6. Layout diagram of the system as modeled. Upper panel: Optical components of the spectrograph. The rays shown emerge at NA = 0.11. The 'slit' extends out of the page. The inset shows an enlarged side view of the Amici lens, with rays emerging from sources along the upper half of the pseudo-slit. Lower panel: overall view of the system, showing the beams reaching the detector plane at the right-hand side. Beams are shown for wavelengths 486.25, 493.125, 500, 506.875 and 513.75 nm.

## 5.1 Waveguide slit and collimator

We again assume that the waveguides or fibers radiate a Gaussian beam of NA = 0.11. Thus the collimator remains the most difficult part of the design since it is a lot faster than most spectrograph collimators, yet the slit should be as long as possible, and the diffraction-limited performance must be preserved.

Figure 6 shows the optical system. The collimator begins with an Amici lens, which is a spherical surface at a particular distance from the source (dependent on the refractive index of the glass at the selected wavelength). This lens begins the collimation of the beam and has the useful property that it is free of spherical aberration of all orders, and both coma and astigmatism at third order. It converts the f/4.5 beam (NA=0.11) to f/10.5, which is much more readily dealt with by the rest of the collimator, and it does so with minimal aberrations. It does have quite strong field curvature, but that is dealt with by curving the input surface from which the waveguides inject the light. This is the reason for the curvature of the injection surface in Figure 6. The pseudo-slit has a length of 6mm (*i.e.* ±3mm).

The two achromats which complete the collimator are essentially arranged as a Petzval lens. They are achromatic doublets with surface curvatures optimised for best image quality.

Overall, the collimator system resembles a medium-power microscope objective layout. The optics may appear oversized in Figure 1, but the size has to allow for the 6 mm length of the slit (out of the page) and for the camera there is the wavelength dispersion at the grating. It is also very important not to overly truncate the Gaussian beams, and the rays in Figure 6 show only the extent of NA = 0.11 *i.e.* the $1/e^2$ points.

The focal lengths were selected to produce a collimated beam diameter of ~22 mm, which as shown in Section 4.1 will produce a spectral resolving power in the vicinity of 50,000 with reasonable grating deviation angles.

The feed into an Amici lens appears ideally suited to use of integrated optics waveguides etched into a block, rather than fibers laboriously arranged into a line using a V-groove template. We assume that the curvature of the 'slit' could be achieved simply by ending the waveguides within the block at the appropriate places, and that the Amici lens would be formed by imposing a spherical surface on the end of the integrated optics block. In other words, the left-most surface in Figure 6 would not exist as an air-glass interface but would instead lie within the integrated optics block and would represent only the curve along which waveguides coming from the left would be terminated.

## 5.2 Diffraction grating

We have assumed the same ideal grating as in Section 4, *i.e.* a transmission grating of 2071 lines/mm. In practice this would be a VPH grating sandwiched between glass plates and minor adjustment of the collimator and/or camera would be required to allow for the effects of the glass plates. Grating quality must be good, to maintain diffraction-limited performance.

## 5.3 Camera

In this example we do not follow version 1 in assuming a detector with arbitrarily small pixels is available. Instead we examine the implications of using CCDs with 15 μm pixels, typical of currently available devices. This necessarily means that the camera must have a long focal length in order to magnify the diffraction-limited images sufficiently for them to be adequately sampled by the pixels. The focal length was chosen to give sampling of about 2.5 pixels per FWHM of the diffraction pattern. As a result the overall dimensions of the instrument are larger than in version 1. It is now 1.9 m from the third camera lens element to the detector but if necessary this could be reduced through use of a telephoto configuration for the camera optics, and/or folded using one or more flat mirrors. In the setting of a large or extremely large telescope, there would be no problem accommodating the linear dimension of ~2m: units could be stacked on shelves or in drawers. The overall size would be substantially reduced through use of detectors with smaller pixels, as are now becoming available.

The long focus camera has the advantage in terms of optical design that it is very slow - with a focal length of 1967 mm and an input beam diameter of 22 mm it is *f*/89 (but this is to the $1/e^2$ points - it is *f*/68 when including the Gaussian beam out to *T*=0.75). The camera also has to deal with beams coming in at a significant range of angles as a result of the dispersion of wavelengths at the grating. But overall it is still an easy design and the triplet shown gives good results, well within the diffraction limit at all field positions and all wavelengths.

This design assumes 8000 pixels in both directions on the detector, *i.e.* a total detector size of 120 × 120 mm, which would be achieved using a 2×2 array of 4K×4K CCDs. The long focal length camera is again helpful in that it can feed such a large focal plane array without any extreme angles.

## 5.4 System performance

Figure 7 shows the spot diagrams over the full range of wavelengths and field (slit) positions. All the geometrical images are well inside the rings representing the notional Airy ring radius that would result from uniform illumination (ZEMAX 'top-hat' profile) across a beam of NA = 0.11. Relative to the Airy ring size, the image quality is markedly superior to that of Figure 4 (version 1). RMS radii vary from about 4 to 7 μm, with just the two extreme ±3mm fields at λ513.75 nm having values up to 12 μm.

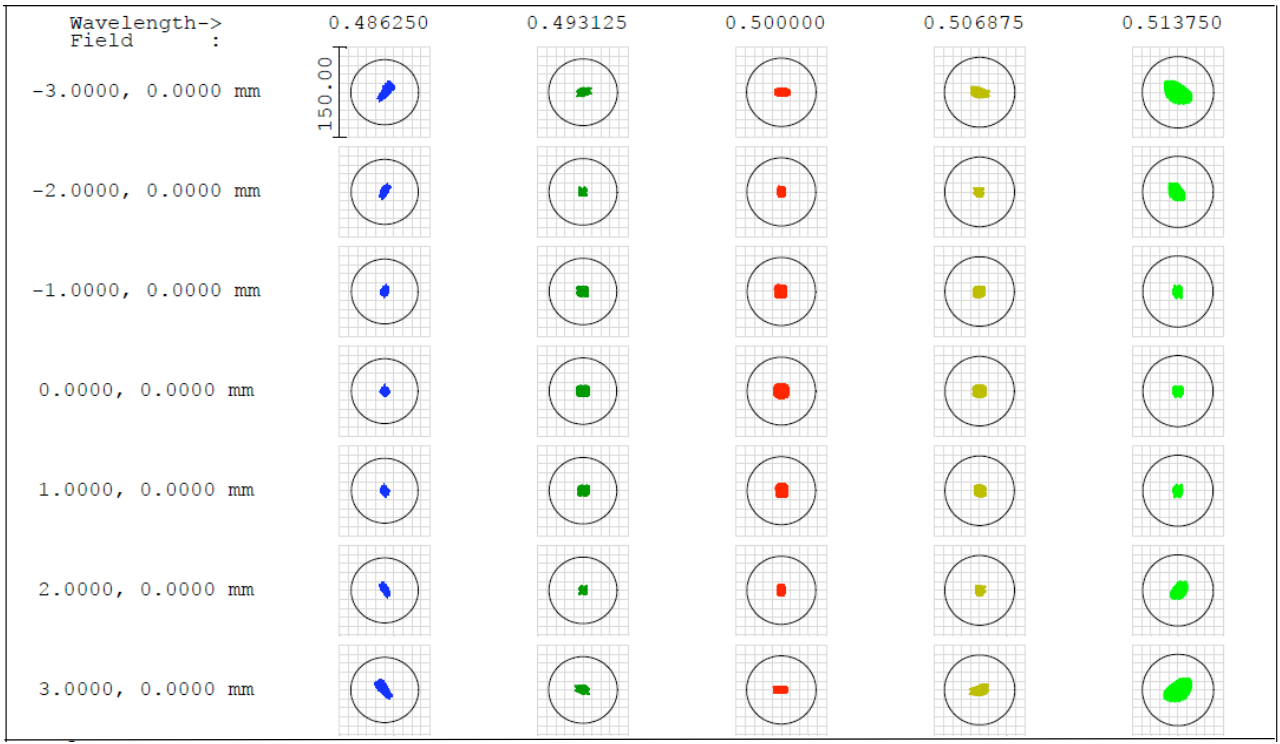

Figure 7. Spot diagrams from λ 486.25 nm to 513.75 nm and for feeds from -3mm to +3 mm along the pseudo-slit. The circles show the Airy ring radius, but diffraction is not taken into account in the spot diagrams. The box size is 150 μm. Illumination is NA=0.11 'top-hat'.

With this good image quality, the system is expected to deliver high and nearly constant spectral resolution. Figure 8 shows the wavelength and spatial profiles, computed using ZEMAX Physical Optics Propagation. The low aberrations result in clean Gaussian profiles in all cases, with little variation of PSF from best to worst despite the much longer slit and wider wavelength range as compared with version 1.

The typical profile is just resolved from a neighboring profile when the two are separated by 42.6 μm on the CCD. The linear dispersion is 4.87 mm on the CCD per nm wavelength shift, so the resolution element is Δλ = 0.00895 nm and resolving power $R$ = 55,900. The worst profile is just resolved from the adjacent resolution element when the separation is 47.9 μm on the CCD, corresponding to Δλ = 0.00984 nm and $R$ = 52,200.

For comparison, the theoretical resolving power can be evaluated from eqn (11) with $\theta_i$ = 31.2°, $d_{1/e2}$ = 21.10 mm and $\lambda_c$ = 500 nm. This gives $R$ = 60,800. The fact that the typical profile's resolution is 7.5% below the theoretical value shows that this design comes very close to the diffraction limit. Even the worst profile has a resolving power only 14% less than the theoretical value. It is clear that truncation of the Gaussian beam is minimal, *i.e.* $T$ is smaller than 0.75, since that value would cause the beam to be broadened by 20% (Sec. 2.2). Nevertheless, some further study of the reasons for the small loss of resolution relative to the diffraction limit is warranted, because the performance of the best profile is similar to that of the typical one, *i.e.* it is not obvious that aberrations are the sole cause of the resolving power loss.

The typical and worst spatial profiles are also shown in Figure 8. The magnification from the waveguide slit to the CCD varies from 20.5 to 20.8. This results in the images of adjacent cores being 38 μm or 2.5 CCD pixels apart. As Figure 8 shows, the profiles are well separated even in the worst case, and so could be separately extracted. This would be optimum in terms of signal/noise since optimally weighted spectral extraction could be used. But it would actually be desirable to pack the spectra more densely in the spatial direction, since many single-mode cores arise from the same multi-mode fiber and there is thus no reason they need to be kept apart.

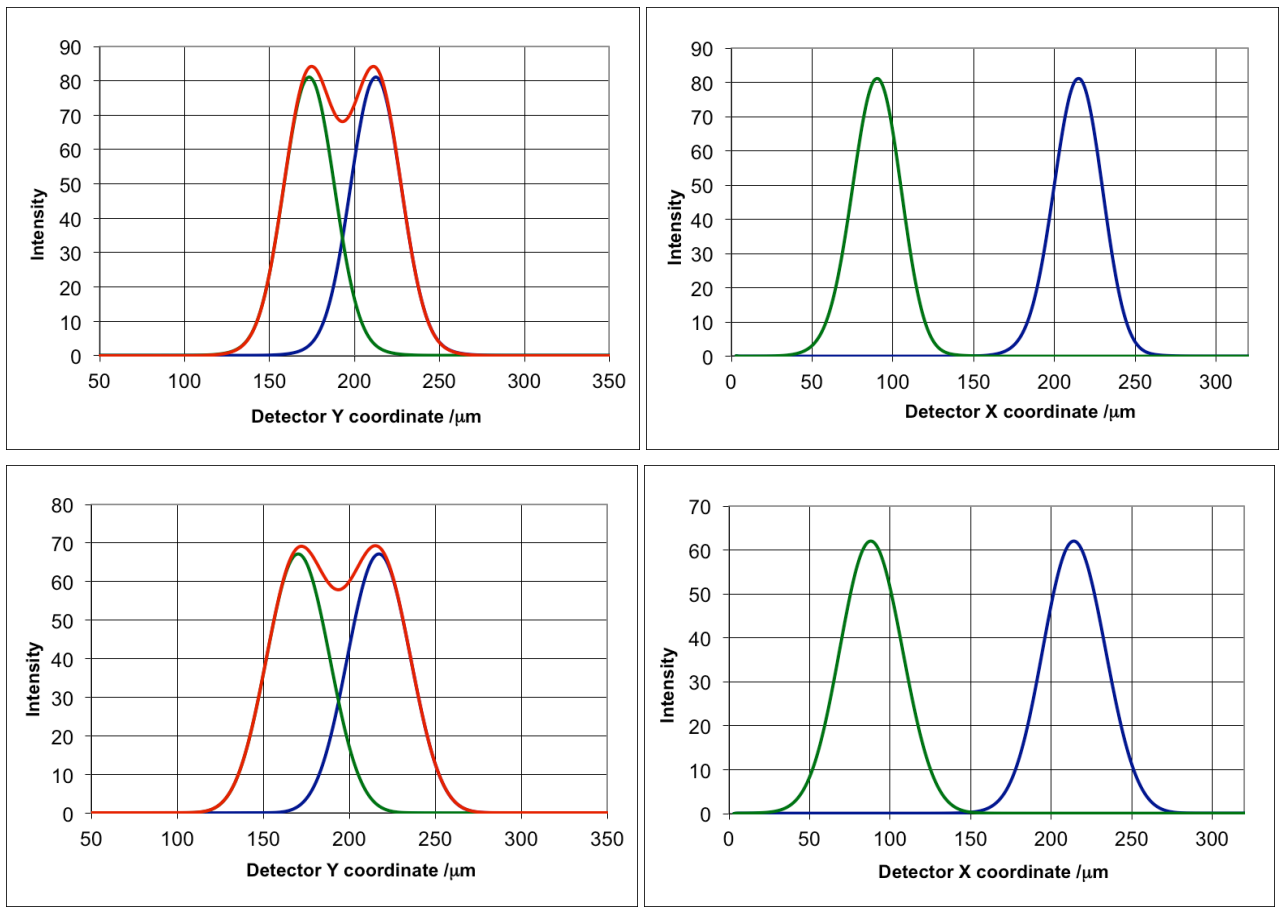

Figure 8. Projected profiles of the images on the detector, accounting for both diffraction and aberrations. Left-hand column: profiles integrated over the spatial direction and displayed along the wavelength axis. The same profile is shown twice, with an offset such that the two are at the resolution limit. The sum is shown in red. Right-hand column: profiles integrated over the wavelength direction and displayed along the spatial direction. Two profiles are shown, corresponding to two source waveguides separated by 6 μm. Top row: the typical image quality (λ 500 nm, slit position +1mm). Bottom row: worst image quality (λ 513.75 nm, -3 mm off axis). The separations of the two profiles along the wavelength axis are 42.6 μm (top left) and 47.9 μm (bottom left).

## 5.5 Version 2 - review

By allowing the system to use relatively simple custom optics instead of restricting it to catalog lenses, the performance has been dramatically improved. Compared with the catalog lens system of version 1, the slit length has been tripled, the wavelength range increased by a factor of 2.2 and the image quality markedly improved. Version 2 feeds a 2×2 array of 4K×4K CCDs and gives good quality spectra at a nearly diffraction-limited resolving power of ~56,000 almost everywhere on the detector.

It covers wavelengths from 486.25 to 513.75 nm (*i.e.* a range of 27.5 nm), with near-optimum sampling of 2.4 - 2.5 CCD pixels (15 μm each) per FWHM. The 8000 pixels in the wavelength direction provide ~ 3,200 independent spectral resolution elements.

The system has been optimised for this wavelength range, and while some limited tuning in wavelength is likely to be possible, this is at present an instrument dedicated to a particular range – similar in principle to the HERMES spectrograph[6] intended for Galactic archaeology studies. Changing wavelength with a transmission grating also requires either the camera or the collimator to be articulated.

The slit length of 6 mm would allow 1000 waveguide cores if they could be spaced 6 μm apart. This slit length is also imaged to a total length of 8000 × 15 μm pixels.

# 6. CONCLUSION AND OUTLOOK

Previous work has shown the application of photonic techniques to astronomical spectrographs[9] has much to offer. Here we have shown that a simple and compact spectrograph design can deliver high resolving power from any telescope or laboratory instrument whose images can be directed into single-mode waveguides. The same design principles can also be applied for observations in the IR[7]. As well as allowing spectrographs to be compact, the use of single-mode fibers provides the opportunity to make enormous signal/noise gains in regions of the spectrum affected by atmospheric OH emission, since this can now be suppressed in SMFs to a high level[5]. We also expect that in future it will be possible to design cameras that are much more compact, as detectors with smaller pixels become available.

The number of waveguide modes in a fiber or waveguide is proportional to $a^2$ where $a$ is the guide core radius, and inversely proportional to $\lambda^2$. Thus seeing-limited images (which are somewhat *larger* in the visible than the IR) result in a very large mode count and hence large spectrograph multiplicity in the visible range. In this sense the visible light spectrographs given here represent the extreme case of the PIMMS#0 concept. Spectrograph count will be reduced by observing in IR bands, and/or by full or partial adaptive optics (AO). One attractive use would be as the spectrograph for a system delivering partially corrected (low Strehl ratio) AO images, since these may be efficiently coupled into few-mode fibres[8]. Compact instruments for diffraction-limited space-borne telescopes constitute another application of the concept.

Topics for further investigation include the use of anamorphic optics to compress the spectra in the spatial direction in order to fit more spectra from individual waveguide cores into a given detector height. There is no need to keep spectra separate, since adjacent spectra have in most cases been derived from the same multi-mode fiber and hence the same source spectrum. In the case of very slow cameras such as Version 2 above, it would even be possible to superimpose the images of a number of cores at each location on the detector, up to the limit imposed by a feasible camera speed (say $f$/2 or $f$/3) in the spatial direction.

# ACKNOWLEDGMENTS

We acknowledge the support of the University of Sydney. Joss Bland-Hawthorn acknowledges the Australian Research Council for the award of a Federation Fellowship.